\documentclass[prl,twocolumn,showpacs,showkeys,preprintnumbers,amsmath,amssymb]{revtex4}

\usepackage{graphicx}
\usepackage{dcolumn}
\usepackage{bm}
\usepackage{amsmath}
\usepackage{amsfonts}
\usepackage{amssymb}
\usepackage{amsthm}
\usepackage{newlfont}
\usepackage{graphicx}
\usepackage{amscd}
\usepackage[all]{xy}

\begin{document}

\title{An explanation of the shape of the universal curve of the Scaling Law \\
for the Earthquake Recurrence Time Distributions
}

\author{Mariusz Bia{\l}ecki}
\email{bialecki@igf.edu.pl}

\affiliation{%
Institute of Geophysics, Polish Academy of Sciences \\
ul.~Ks.~Janusza 64, 01-452 Warszawa, Poland}%

\date{October 26, 2012}%

\begin{abstract}
This paper presents an explanation of a possible mechanism underlying the shape of the universal curve of 
Scaling Law for Earthquake Recurrence Time Distributions. The presented simple stochastic cellular automaton model 
is reproducing the gamma distribution fit with the proper value of the parameter $\gamma$ characterizing Earth's seismicity and also imitates a deviation from the fit at the short interevent times, as observed in real data.   

Thus the model suggests an explanation of the universal pattern of rescaled Earthquake Recurrence Time Distributions in terms of combinatorial rules for accumulation and abrupt release of seismic energy. 
\end{abstract}


\maketitle


Analyzing seismic catalogs, Corral \cite{CorralSL} has determined that the probability densities of the waiting times between earthquakes 
for different spatial areas and magnitude ranges can be described by a unique universal distribution if the time is rescaled with the mean rate of occurrence. 

To unify diverse observations the spatiotemporal analysis was carried out as follows. 
Seismicity is considered as result of a dynamical process in which collective properties are largely independent on the physics of the individual earthquakes.
Following  \cite{BakUSLE}, events are not separated into different kinds (foreshocks, mainshocks, aftershocks) nor the crust is divided into provinces with different tectonic properties. 
Then, a region of the Earth is selected, as well as temporal period and a minimum magnitude $M_c$ (for conditions and other details see \cite{CorralSL, CorralLNP}). 
Events in this space-time-magnitude window are considered as a point process in time (disregarding the magnitude and the spatial degrees of freedom) and are characterized only by its occurrence time $t_i$, with $i=1 \ldots N(M_c)$. 
Then the recurrence (or waiting) time $\tau_i$  is defined by $\tau_i = t_i - t_{i-1}$. 

The entire Earth has been analyzed by this method and it appears that different regions' probability densities of waiting times, 
rescaled by the mean seismic rate, as a function of the rescaled recurrence time collapse onto a single curve $f$  \cite{CorralSL}:
\begin{equation}
D(\tau; M_c) = R(M_c) f(R(M_c)\tau),
\label{eq:ScLaw}
\end{equation}
where  mean seismic rate $R(M_c)$ is given by $R(M_c)={N(M_c)}/{T}$, (here $T$ is a total time into consideration), 
and recurrence-time probability density $D(\tau;M_c)$ is defined as
$D(\tau;M_c)={Prob [ \tau < \text{recurrence time} < \tau+ d\tau ] }/{\tau}$.
The so called scaling function $f$ admits a fit in the form of a generalized gamma distribution
\begin{equation}
f^{fit}(\theta)=C \theta^{\gamma-1}\exp(-\frac{\theta^\delta}{\beta}),
\label{eq:funcf}
\end{equation}
where $\gamma=0.67\pm0.005$, $\beta=1.58\pm0.15$, $\delta=0.98\pm0.05$, $C=0.5\pm0.1$, and $\theta=R\tau$ is dimensionless recurrence time. The value of $\delta$ can be approximated to $1$.  
The present characterization of the stochastic spatiotemporal occurrence of earthquakes by means of a unique law would 
indicate the existence of universal mechanisms in the earthquake-generation process \cite{CorralSL}.

This paper presents an  explanation of a possible mechanism underlying the shape of the universal curve in terms of a cellular automaton model called Random Domino Automaton (RDA). 
The simple rules for evolution of the model, being a slowly driven system, rely on accumulation and abrupt release of energy only, which fit well to the described above procedure of neglecting individual properties of earthquakes.
We show that RDA reproduces the rescaled distribution of recurrence times.

As can be seen from the original work \cite{CorralSL} as well as from further studies \cite{Marekova},
results obtained from various earthquake catalogs show a deviation from the gamma distribution at the short 
interevent times. This holds from worldwide to local scales and for quite different tectonic environments.

It is remarkable that the presented model reproduces  also this deviation. 
Thus the model suggests an explanation of the universal pattern of rescaled Earthquake Recurrence Time Distributions in terms of its combinatorial rules for accumulation and release of seismic energy. 

So far, some insight of the origin of the gamma distribution as well as examination the recurrence statistics of a range of cellular automaton earthquake models are presented in \cite{WeathRecCA}. It is shown there, that only one model, the Olami-Feder-Christensen automaton, has recurrence statistics consistent with regional seismicity for a certain range of conservation parameter of that model.

The Random Domino Automaton (RDA) was introduced as a toy model of earthquakes \cite{BiaCz-Mon, BiaCzAA, BiaMotzkin, BiaFRDA}, 
but  can be also regarded as an extension of well known
1-D forest-fire model proposed by Drossel and Schwabl \cite{DSfire}.
As a field of application of RDA we have already studied its relation to Ito equation \cite{CzBia-Mon, CzBiaTL, CzBiaAG} and
to integer sequences \cite{BiaMotzkin}. 
We point out also other cellular automata models \cite{PachMin, Pach08, Pach09, Pach10} giving an insight into diverse specific 
aspects of seismicity, including predictions.

\begin{table}[t]%
\caption{States of RDA for the size of the lattice $N=5$.}
\label{tab:N5}
\begin{ruledtabular}
\begin{tabular}{ccc}
state label &  example & multiplicity  \\
\hline\hline
1 & $ \hookrightarrow | \ \ \ | \ \ \ | \ \ \ | \ \ \ | \ \ \ | \hookleftarrow $	& 1 \\
\hline
2 & $ \hookrightarrow | \ \ \ | \ \ \ | \ \ \ | \ \ \ | \bullet | \hookleftarrow $	& 5  \\
\hline
3 & $ \hookrightarrow | \ \ \ | \ \ \ | \ \ \ | \bullet | \bullet | \hookleftarrow $	& 5 \\
\hline
4 & $ \hookrightarrow | \ \ \ | \ \ \ | \bullet | \ \ \ | \bullet | \hookleftarrow $	& 5 \\
\hline
5 & $ \hookrightarrow | \ \ \ | \ \ \ | \bullet | \bullet | \bullet | \hookleftarrow $	& 5 \\
\hline
6& $ \hookrightarrow | \ \ \ | \bullet | \ \ \ | \bullet | \bullet | \hookleftarrow $	& 5 \\
\hline
7 & $ \hookrightarrow | \ \ \ | \bullet | \bullet | \bullet | \bullet | \hookleftarrow $& 5  \\
\hline
8 & $ \hookrightarrow | \bullet | \bullet | \bullet | \bullet | \bullet | \hookleftarrow $ & 1 \\
\end{tabular}
\end{ruledtabular}
\end{table}

The RDA is characterized as follows:\\
- space is 1-dimensional and  consists of $N$ cells; periodic boundary conditions are assumed;\\
- cell may be empty or occupied by a single ball;\\
- time is discrete and in each time step an incoming ball hits one arbitrarily chosen cell (the same probability for each one). 
The balls are interpreted as energy portions.
   
The state of the automaton changes according to the following rule:\\
$\bullet$ if the chosen cell is empty it becomes occupied with probability $\nu$; with probability $(1-\nu)$ the incoming ball is rebounded and the state remains unchanged; \\
$\bullet$ if the chosen cell is occupied, the incoming ball provokes an avalanche with probability $\mu$ (it removes balls from hit cell and from all adjacent cells); with probability $(1-\mu)$ the incoming ball is rebounded and the state remains unchanged.

The parameter $\nu$ is assumed to be constant but the parameter $\mu=\mu_i$ is allowed to be a function of size $i$ of the hit cluster. This extension with respect to Drossel-Schwabl model leads to substantial novel properties of the automaton.
Note, only ratio of these parameters $\mu_i/\nu$  affects properties of the automaton --  changing of $\mu$ and $\nu$ proportionally corresponds to a rescaling of time unit. 

The remarkable feature of the RDA  is the explicit one-to-one relation between details of the dynamical rules of the automaton
(represented by rebound parameters $\mu_i/\nu$) 
and the produced stationary distribution $n_i$ of clusters of size $i$, which implies distribution of avalanches $w_i$. 
It shows how to reconstruct details of the "macroscopic" behavior of the system from simple rules of 
"microscopic" dynamics. 

Various sizes $N$ of RDA can be considered in order to explain the shape of the universal curve of Scaling Law.
It appears results for quite a small size $N=5$ are enough to explain the idea and allow to keep the reasoning concise and detailed. RDA for a bigger size of the lattice behaves similar-like and the overall picture remains the same, as  
results from explanations given below.

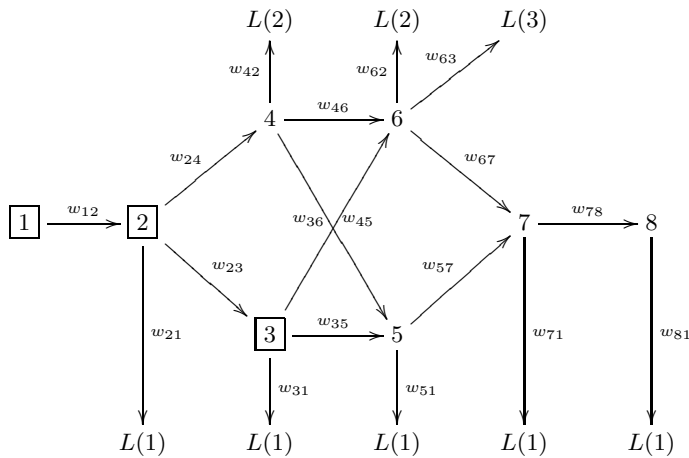
\begin{figure}[t]  
\begin{displaymath}
\xymatrix{  && L(2) & L(2) & L(3)& \\
&&4\ar[u]^{w_{42}}\ar[r]^{w_{46}}\ar[rdd]^{w_{45}}&6\ar[u]^{w_{62}}\ar[ur]^{w_{63}}\ar[dr]^{w_{67}}&& \\
\boxed{1}\ar[r]^{w_{12}}&\boxed{2}\ar[rd]^{w_{23}}\ar[ru]^{w_{24}}\ar[dd]^{w_{21}}&&&7\ar[r]^{w_{78}}\ar[dd]^{w_{71}}&8 \ar[dd]^{w_{81}}\\
&&\boxed{3}\ar[r]^{w_{35}}\ar[uur]^{w_{36}}\ar[d]^{w_{31}}&5\ar[d]^{w_{51}}\ar[ur]^{w_{57}}&& \\
&L(1)&L(1)&L(1)&L(1)&L(1)
}
\end{displaymath}
\caption{A state diagram for RDA of size $N=5$. Arrows with respective weights indicate possible transitions; those with avalanches are ended with symbol "L". A state is boxed, if it is possible to get it directly after an avalanche.   
 } \label{fig:Fig1}
\end{figure}

RDA is also a Markov chain \cite{BiaFRDA}.
It is convenient to define states $i$ up to translational equivalence.
Thus, in example, for $N=5$, instead of $2^5$, there are $8$ states only -- see Table \ref{tab:N5}. 
Such space of states is irreducible, aperiodic and recurrent.
Transition matrix $\mathbf{p}$, where 
$[ \mathbf{p} ]_{ij} = \text{probability of transition} \quad i \longrightarrow j$,
for $N=5$ is of the form
\begin{widetext}
\begin{equation}
\mathbf{p} = \frac{1}{5}
\left( \begin{array}{cccccccc}
 5-5\nu &   5\nu&       0&           0&               0&            0&               0&          0 \\ 
 \mu_1  &   5-\mu_1-4\nu & 2\nu       &   2\nu&            0&            0&               0&          0\\
 2\mu_2 &    0&          5-2\mu_2-3\nu& 0&               2\nu&         \nu&              0&          0\\
   0    &        2\mu_1&      0&            5-2\mu_1-3\nu&  \nu&            2\nu&            0&          0\\
  3\mu_3&    0&          0&            0&               5-3\mu_3-2\nu& 0&               2\nu&       0\\
     0  &        2\mu_2&      \mu_1&          0&               0&            5-2\mu_2-\mu_1-2\nu& 2\nu&        0\\
  4\mu_4&    0&          0&            0&               0&            0&               5-4\mu_4-\nu& \nu\\
  5\mu_5&    0&          0&            0&               0&            0&               0&          5-5\mu_5 
\end{array} \right)
\label{eq:transM5}
\end{equation}
\end{widetext}
Stationary distribution is given by
\begin{equation}
v \cdot \mathbf{p}=v.
\label{eq:MP}
\end{equation}
The evolution of the system is represented in Figure \ref{fig:Fig1}. Arrows between states $i$ and $j$, with respective weights $w_{ij}$, indicate possible transitions. A symbol $L(j)$ depict an avalanche to state $j$. The density of the system is growing from left side (state $1$ has density $\rho=0$) to right side (up to density $\rho=1$ for state $8$).

The expected time between two consecutive avalanches $T_{av}$ may be expressed by various formulas \cite{BiaFRDA}.
For example 
\begin{equation}
T_{av}= \frac{\left\langle w \right\rangle + 1}{1-P_r}, 
\label{eq:tavav}
\end{equation}
where $\left\langle w \right\rangle$ is the average avalanche size and $P_r$ is the probability that the incoming ball is rebounded both form empty or occupied cell .

The probabilities $v_i$ of states $i$ obtained from condition \eqref{eq:MP} allow to determine the distribution of frequency $f_i$ of avalanche of size $i$, if rebound parameters $\mu_i/\nu$ are given.
There exists also a systematic procedure of obtaining approximate values of rebound parameters $\mu_i/\nu$, which produce requested 
distribution of avalanches \cite{BiaFRDA1to1}. The approximation comes from nonexistence of exact equations for (stationary) distribution of clusters $n_i$  for sizes bigger than $4$ (see \cite{BiaFRDA}).
We have used this procedure to obtain values $\mu_i/\nu$ that give noncumulative inverse-power distribution of avalanches presented in Table \ref{tab:munu}.

To calculate the distribution of waiting times, each path starting from a state reached after an avalanche and ending 
with an avalanche is considered. There are $42$ such paths for size $N=5$. 
Each path is assigned its respective probabilities, that a total passage time is equal to $1,2,\ldots$ time steps.  
 
Respective weights $S_k$ describing how often the system starts from initial state $k$  are given by
\begin{equation}
S_k = \frac{\sum_{i > k} v_i p_{ik} }{\sum_k \sum_{i > k} v_i p_{ik}}.
\label{eq:pfi}
\end{equation}
For $N=5$ initial states are $1,2$ and $3$.

The expected time of stay in a state $k$ is 
\begin{equation}
t_{av}(k)= \sum_i  p_{kk}^{(i-1)} (1-p_{kk}) \cdot i  = (1-p_{kk})^{-1}.
\label{eq:tavk}
\end{equation}

Probability of stay in given state $k$ for a time equal to $i$ time steps is given by 
\begin{equation}
T^k_i=p_{kk}^{(i-1)} (1-p_{kk}),
\label{eq:tk}
\end{equation}
and all possible values are aggregated in a vector $T^k$ with $i$-th component equal to $T^k_i$.
For path through two consecutive states $k$ and $l$, respective probability of time of stay in both of them equal to
$j$ time steps is defined by 
\begin{equation}
T^{kl}_j=   (T^k \star T^l)_j = \sum_{n=1}^{j-1}   T^k_j T^l_{n-j}.
\label{eq:prodT}
\end{equation}
For a path through three states $k,l,m$ we have $T^{klm}=   (T^k \star T^l)\star T^m$, and so on for longer paths. 

The probability rates $w_{ij}$ for transition   $i \rightarrow j$ where $i \neq j$ are just normalized probabilities $p_{ij}$, 
namely
\begin{equation}
w_{ij}=\frac{p_{ij}}{\sum_{j \neq i} p_{ij}}.
\label{eq:}
\end{equation}

Thus for a path $i_1,i_2,\ldots,i_{k-1},i_k$ there is assigned a combined weight 
\begin{equation}
W^{i_1 i_2 \ldots i_{k-1} i_k}=S_{i_1} \cdot w_{i_1 i_2} \cdot \ldots \cdot w_{i_{k-1} i_k},
\label{eq:weightW}
\end{equation}
as well as combined weigted time vector
\begin{equation}
\Omega^{i_1 i_2 \ldots i_k}= W^{i_1  i_2 \ldots i_k}\cdot T^{i_1 i_2 \ldots i_k}.
\label{eq:WTvec}
\end{equation} 
The $i$th component of the  vector $\Omega^{i_1 i_2 \ldots i_k}$ gives a contribution to waiting time equal to $i$ 
coming from a path $i_1,i_2,\ldots,i_k$.
Summing up those vectors for all possible paths we end with a distribution of waiting times. One can 
obtain also a distribution related to avalanches of chosen size. 
For example, if such sum is made for  paths related to avalanches of size $2,3,4$ and $5$ only,  a distribution of waiting times related to avalanches of size bigger then $1$ is obtained. 

Rebound parameters presented in Table \ref{tab:munu} were chosen in order to obtain noncumulative distribution of avalanches in the form consistent with Gutenberg-Richter law. 
The exact value of power (here $2.1$) does not affect results of the construction substantially.

\begin{table}[t]%
\caption{Approximate values of rebound parameters ${\mu_i}$ and respective avalanche distribution  $w_i$
The parameter $\nu=0.25$.}
\label{tab:munu}
\begin{ruledtabular}
\begin{tabular}{l|ccccc}
$i$& $1$ & $2$ & $3$ & $4$ & $5$ \\ 
\hline
$\mu_i$ & 0.999060 & 0.388232 & 0.284504 & 0.097650 & 0.045810 \\
$w_i$   & 0.413247 & 0.102851 & 0.042587 & 0.022351 & 0.014306
\end{tabular}
\end{ruledtabular}
\end{table}

The system has average density $\rho=0.273885$,  average avalanche size  $\left\langle w \right\rangle= 1.52458$
and average time between avalanches $T_{av}= 21.2027$.
The parameter $P_r= 0.880932$ shows, that most of incoming balls are rebounded. 
Expected times of staying in all states are presented in Table \ref{tab:tav}.
Great majority of avalanches leads to empty state ($S_1=0.755449$), roughly every fifth avalanche leads to state $2$ ($S_2=0.205253$), and roughly every twenty fifth to state $3$ ($S_3=0.0392984$).

\begin{table}[t]%
\caption{Approximate expected stay times in states for the size of the lattice $N=5$.}
\label{tab:tav}
\begin{ruledtabular}
\begin{tabular}{l|cccccccc}
 state     & 1   & 2   & 3   & 4  & 5  & 6  &7    & 8 \\
 \hline
$t_{av}$& 4.0 & 2.5 & 3.3 & 1.8& 3.7& 2.2& 7.8 & 21.8
\end{tabular}
\end{ruledtabular}
\end{table}

\begin{figure}[t]  
	\centering
	\includegraphics[width=8cm]{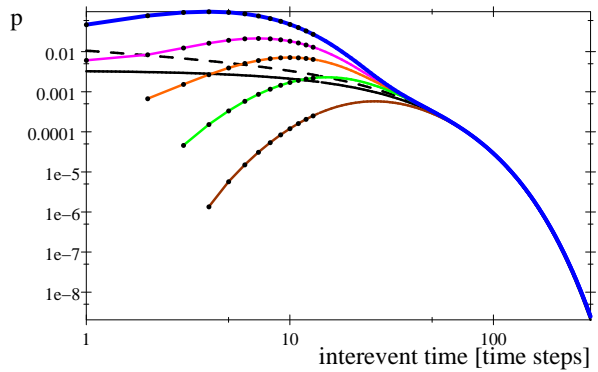} 
	\caption{ A plot of distributions of interevent times for RDA with N=5. The upper line is for avalanches of all sizes, 
	the curve below is for avalanches of size bigger then $1$, the next is for sizes bigger then $2$, and so on. The lowest curve
	counts only avalanches of size $5$. 
	The dashed line is a plot of fitted gamma distribution $ay^{(\gamma-1)}e^{-\frac{y}{b}}$; the solid line below is a plot of fitted exponential curve $ a' e^{-\frac{y}{ b'}}$. }
	\label{fig:N5lin2}
\end{figure}

Figure \ref{fig:N5lin2} presents obtained distributions of waiting times up to $300$ time steps. 
The upper curve is for avalanches of all sizes, the next is for avalanches of size bigger then $1$, and so on.  
The lowest curve count avalanches of size $5$ only.

Dashed line is a fitted gamma distribution $ay^{(\gamma-1)}e^{-\frac{y}{b}}$, where  $\gamma=0.67$,  $b=22.4$ and  
$c=0.011$.  This fit is done for  points with time coordinate from $60$ ($\chi^2=5.5346\cdot 10^{-11}$).
Values of the parameters $b$ and $c$ can be rescaled, depending on their relation to physical quantities (time, number of earthquakes).
The parameter $\gamma$ is a fixed parameter, with exactly the same value which characterize Earth's seismicity.
The solid line below is a plot of fitted exponential curve $ a' e^{-\frac{y}{ b'}}$. 

Thus, the exponential part of the universal curve comes from distributions of biggest avalanches. 
In the presented example the biggest $t_{av}$ is for the state $8$ containing single cluster of size $5$ (see Table \ref{tab:tav}).
Thus its contribution to the overall waiting time distribution dominates for bigger times (compare formulas \eqref{eq:tavk} and \eqref{eq:tk}). Also state $7$ containing single cluster of size $4$ contribute, but it is decaying more rapidly.

The other part of the universal curve, comes from  contributions of avalanches of smaller sizes.
Its shape is a result of composition of many possible paths of the evolution, depicted in Figure \ref{fig:Fig1}.
For bigger sizes $N$ there are much more possible paths (i.e. $1554$  for $N=7$) through states containing many clusters with 
comparable times $t_{av}$. Their composition flatten the curve. 
Moreover, calculation shows this effect produces a surplus (comparing to the gamma fit) for small waiting times,
which is evident in real earthquakes data \cite{CorralSL, Marekova}. The size of the surplus can be reduced by 
omitting of a contribution of smallest avalanches (also not recorded in real data).  

Note, that due to the incompleteness of the seismic catalogs in the short-time scale, usually real data are not displayed on  plots for very short time intervals. Thus, the obtained theoretical curve, shown in Figure \ref{fig:Fig1}, may be similarly cut for small times. If it is done for time, say, smaller then $10$, it reflects the shape of real data.   

Thus, the presented model suggests that the origin of a universal curve is of combinatorial nature of accumulation and abrupt release of energy according to the rules depending on some parameters defining probabilities dependent on size of 
energy portions, as described above.

\section*{Acknowledgement}
The author would like to express his gratitude to Professor Zbigniew Czechowski
for helpful discussions.


\end{document}